\documentclass[11pt,a4paper,table,xcdraw]{paper}
\usepackage[utf8]{inputenc}
\usepackage[left=2.5cm,right=2.5cm,top=2.5cm,bottom=2.5cm]{geometry}
\usepackage{authblk}
\usepackage{graphicx}
\usepackage{latexsym}
\usepackage{amsmath}
\usepackage{amssymb}
\usepackage{tikz}
\usetikzlibrary{shapes.geometric, arrows, quantikz, backgrounds}
\usepackage{exceltex}
\usepackage{array}
\usepackage{multirow}
\usepackage{float}
\usepackage{color}
\usepackage{soul}
\usepackage{xcolor}
\usepackage{tcolorbox}
\usepackage{braket}
\usepackage{bbold}
\usepackage{cancel}
\usepackage{makeidx}
\usepackage{imakeidx}
\usepackage{algorithm}
\usepackage{algorithmicx}
\usepackage[noend]{algpseudocode}
\usepackage{indentfirst}
\usepackage{blochsphere}
\usepackage{mathtools, cuted}
\usepackage[title]{appendix}
\usepackage{natbib}
\usepackage{breakcites}
\usepackage{flushend}
\usepackage{hyperref}
\hypersetup{colorlinks,linkcolor={blue},citecolor={blue}}

\newcommand{\bb}[1]{\textbf{#1}}

\tikzstyle{startstop} = [rectangle, rounded corners, minimum width=2cm, minimum height=1cm,text centered, draw=black]
\tikzstyle{process} = [rectangle, minimum width=2cm, minimum height=1cm, text centered, draw=black]
\tikzstyle{decision} = [diamond, minimum width=2cm, minimum height=1cm, text centered, draw=black, aspect=2]
\tikzstyle{arrow} = [thick,->,>=stealth]
\tikzstyle{arrowR} = [thick,->,>=stealth, dashed]
\tikzstyle{coord} = [coordinate]

\begin{document}

\title{\centering Implementable Hybrid Quantum Ant Colony Optimization Algorithm}

\author[1]{{ M. Garcia de Andoin}  \thanks{E-mail: mikelgda@gmail.com}}
\author[2]{{ J. Echanobe}}

\affil[1]{{\small Physical Chemistry Department, Faculty of Sciences and Technology, University of the Basque Country, Leioa, Bizkaia, 48940 Spain}}
\affil[2]{{\small Electricity and Electronics Department, Faculty of Sciences and Technology, University of the Basque Country, Leioa, Bizkaia, 48940 Spain}}

\maketitle
\begin{abstract}
We propose a new hybrid quantum algorithm based on the classical Ant Colony Optimization algorithm to produce approximate solutions for NP-hard problems, in particular optimization problems. First, we discuss some previously proposed Quantum Ant Colony Optimization algorithms, and based on them, we develop an improved algorithm that can be truly implemented on near-term quantum computers. Our iterative algorithm codifies only the information about the pheromones and the exploration parameter in the quantum state, while subrogating the calculation of the numerical result to a classical computer. A new guided exploration strategy is used in order to take advantage of the quantum computation power and generate new possible solutions as a superposition of states. This approach is specially useful to solve constrained optimization problems, where we can implement efficiently the exploration of new paths without having to check the correspondence of a path to a solution before the measurement of the state. As an example of a NP-hard problem, we choose to solve the Quadratic Assignment Problem. The benchmarks made by simulating the noiseless quantum circuit and the experiments made on IBM quantum computers show the validity of the algorithm.
\end{abstract}

\section{Introduction}
\label{sec:intro}
Ant Colony Optimization (ACO) was proposed by A. Colorni and M. Dorigo in the 1990's for solving combinational optimization problems \citep{ACOoriginal}. This bio-inspired algorithm mimics the foraging strategy, in which each individual tries to find the shortest path to the food based on the information of its predecessors. This indirect communication is made by placing pheromones along the path an individual traverses. The pheromones will be stronger the better is the food and the shorter the path to it. This way, the ants will optimize the distance between the food and the colony. Using this metaheuristic, researchers have solved instances of NP-hard combinational problems, many of them oriented to graphs, such as Travelling Salesman Problem (TSP) \citep{ACOoriginal,ACS-TSP}, Vehicle Routing Problem (VRP) \citep{AS-VRP}, Quadratic Assignment Problem (QAP) \citep{AS-QAP} or function optimization \citep{ACO-MIN}.

Quantum computing and Quantum algorithms have been rapidly growing since the first very successful results were published in the early 1990's. Grover's searching algorithm \citep{Grover} or Shor's factoring algorithm \citep{Shor} proved to outperform any other classical algorithm in regard to time complexity. In 1996, A. Narayanan et al. \citet{QIGA} proposed quantum-inspired genetic algorithms, in which mechanisms used in quantum computing were applied to improve evolutionary algorithms. A few years later, \citet{QEA} proposed Quantum Evolutionary Algorithms (QEA), which speedups the classical evolutionary algorithms based on the same principles.

Based on QEA and ACO, \citet{NovelQACO} proposed a quantum inspired ant colony optimization algorithm (QIACO). The main novelty of QIACO is the pheromone representation on the quantum state, using a rotation gate to direct the measurement of the system to the optimal solution. Similarly, P. Li and H. Wang proposed a QIACO algorithm based on the Bloch spherical search (BQIACO) \citep{BQACO}. This algorithm takes advantage of the Bloch sphere representation, applying rotation gates in order to move the state to the optimal solution. Mimicking the random search pattern in ACO, both algorithms implemented forced exploration strategies, using CNOT gates in QIACO and Hadamard gates in BQIACO to shift around the qubits one by one.

However, neither QIACO or BQIACO can be implemented on a real quantum computer. Due to the limitations of the information quantum mechanics allows us to retrieve from a quantum state, the pheromone update strategies that both propose can't be used. In this article, we propose a new quantum version of ACO that is implementable on a quantum computer.

In the section \ref{sec:PreviousWork} we introduce the original QIACO algorithm and a recently proposed QACO algorithm. Section \ref{sec:ImplementableQACO} develops the new algorithm and a discussion on the parameter optimization. Section \ref{sec:Implementation} presents the results obtained solving QAP both by simulating the algorithm and implementing it on the currently available IBM quantum computers \citep{IBM}. Finally, section \ref{sec:Conclusion} discusses conclusions, possible improvements and future work.

There is some confusion in the names, as the original authors often label their algorithms as ``quantum" despite being fully classical. Instead, we refer to this algorithms as ``quantum inspired". To label our algorithm we followed the criterion used by other authors \citep{Hybrid1,QCChemistry}. This is better suited to the fact that the hybrid quantum algorithms take advantage of the quantum mechanical properties of the states, while subrogating some calculations to classical algorithms.

\section{Previous works}
\label{sec:PreviousWork}
The algorithm presented in this article is an improvement based on the work of \citet{NovelQACO}. There, they proposed a quantum approach for the classic Ant Colony Optimization algorithm, in which each state of the computational basis represents a possible solution for the problem.

The information about the pheromones is then coded in the quantum state of the system. To match the qubit and the pheromone representation, they used the Hyper-Cube Framework (HCF) proposed by \citet{HCF}. As the HCF limits the pheromone values to the range [0,1], the probability of measuring the excited state of a qubit can be set to be the same as the probability for an ant of choosing said edge.

To achieve that, they used a rotation gate around the Y-axis on the Bloch sphere. This way, every qubit is assigned an angle, this being $\pi/4$ assuming the initial state of the qubits is $\ket{0}$. From this point on, we suppose that all qubits start always at the state $\ket{0}$. Each time a solution is obtained, it is compared to the best solution so far. The rotation angle for the next generation is then updated using a lookup table. 

As in ACO, the algorithm must allow random exploration of new solutions. This is achieved by generating a random number $0 < p < 1$ for each qubit. If $p$ is greater than the exploration parameter $p_e$, the outcome of the qubit will be random. Else, it will follow the pheromones as in ACO. 

Lately, ACO has been improved and used to solve different problems more efficiently. For example, for automated guided vehicles \citep{AVG}, for topology-based link prediction \citep{Link} and for query optimization \citep{query}. These implementations are based on the parallel nature of quantum systems. The ability of having superposed quantum states that represent different possible solutions, enhance the ability to avoid falling into local minima.

Nevertheless, QIACO is not implementable in a quantum computer. On one hand, in order to use the lookup table, one has to know exactly the state of each qubit. As there is no way to achieve this out of a simulation or repeating the experiment until obtaining the statistics of the distribution of states, the version of QACO we propose in this article uses a slightly different approach to the pheromone update strategy. On the other hand, the exploration strategy of QIACO cannot be implemented with quantum gates. The strategy they proposed takes a measurement of the qubits, consequently destroying the quantum state. 

Recently, a quantum algorithm for ACO has been proposed, the MNDAS algorithm \citep{QforACO}. This algorithm uses $x$ qubits to code all possible paths, $d$ for the pheromones and 3 qubits as registers. The qubits are initialized in a superposition state using Hadamard gates in order to give each path the same weight. Then, the algorithm takes an iterative approach to the problem using an oracle. Its main function is to select $n$ possible paths as in ACO and to update the pheromone trails accordingly. Before ending each iteration, the oracle performs an operation to evaporate the pheromones on the selected trails. The convergence of the algorithm is assured by preventing evaporation to occur on the best path found so far. After a fixed number of iterations, a quantum amplitude amplification procedure is made in order amplify the probability density of the solution, and then measured.

One problem of MNDAS is its lack of implementability in near-term quantum computing systems. The number of qubits and the couplings it employs is far from achievable. The implementation of the algorithm on any currently available quantum computer would need an unaffordable amount of SWAP gates, introducing a large amount of noise in the system. Another problem with this algorithm is the introduction of a highly demanding oracle. The amount of calculations it needs to perform in each iteration makes it difficult for a quantum computer to maintain coherence after all the gates that are applied. In contrast, our algorithm tackles this problems by using an iterative quantum algorithm that doesn't rely on perfectly error corrected qubits. The amount of gates our algorithm performs in each iteration makes is suitable to be implemented in current quantum computers.

Furthermore, the initialization of MNDAS requires to compute the weights of all possible paths. This completely defeats the purpose of using a metaheuristic algorithm. Once the paths weights are obtained, the optimal solution can be obtained using a regular search algorithm $\mathcal{O}(n)$, or a Grover algorithm $\mathcal{O}(n^{1/2})$. In contrast, MNDAS has a complexity of $\mathcal{O}(Kn+n^{1/2})$, with $K$ a constant, so it falls behind already existing solutions.

\section{Proposed implementable QACO algorithm}
\label{sec:ImplementableQACO}
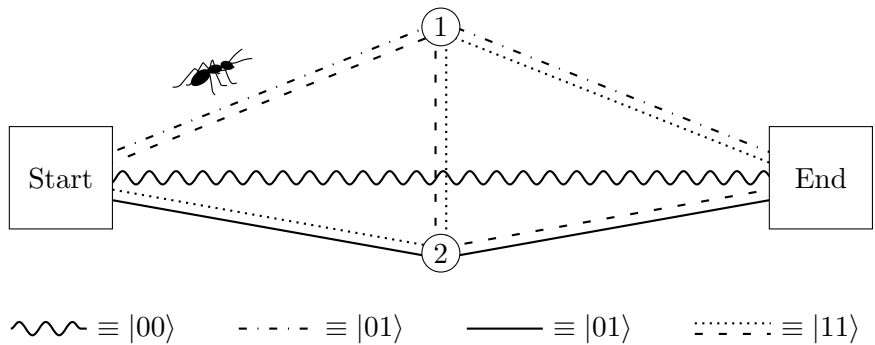
\begin{figure*}[h]

\makebox[\textwidth][c]{
\begin{tikzpicture}[node distance=5cm]
    \node (start) [rectangle, draw=black, fill=white, fit={(0,0) (1.1,1.1)}, label=center:Start] {};
    \node (centro) [right of=start] {};
    \node (nodo2) [circle, draw=black, fill=white, below of=centro, minimum size=0.5cm, yshift=4cm, label=center:2] {};
    \node (nodo1) [circle, draw=black, fill=white, above of=centro, minimum size=0.5cm, yshift=-3cm, label=center:1] {};
    \node (final) [rectangle, draw=black, fill=white, fit={(0,0) (1.1,1.1)}, right of=centro, label=center:End] {};
    \node (a1) [below of=centro, yshift=3cm, xshift=-4cm] {$\equiv\ket{00}$}; 
    \draw [draw=black, style={decorate, decoration=snake}, thick] ([xshift=-1cm]a1.west)--(a1.west);
    \node (a2) [right of=a1, xshift=-2cm] {$\equiv\ket{01}$};
    \draw [draw=black, loosely dashdotted, thick] ([xshift=-1cm]a2.west)--(a2.west);
    \node (a3) [right of=a2, xshift=-2cm] {$\equiv\ket{01}$};
    \draw [draw=black, thick] ([xshift=-1cm]a3.west)--(a3.west);
    \node (a4) [right of=a3, xshift=-2cm] {$\equiv\ket{11}$};
    \draw [draw=black, dotted, thick] ([xshift=-1cm, yshift=2pt]a4.west)--([yshift=2pt]a4.west);
    \draw [draw=black, loosely dashed, thick] ([xshift=-1cm, yshift=-2pt]a4.west)--([yshift=-2pt]a4.west);
    \draw [-,opacity=0] ([xshift=-4cm]start.center) -- ([xshift=4cm]final.center);
    
    \node (hormiga) [right of=start, xshift=-3cm, yshift=1.4cm] {\includegraphics[width=30pt]{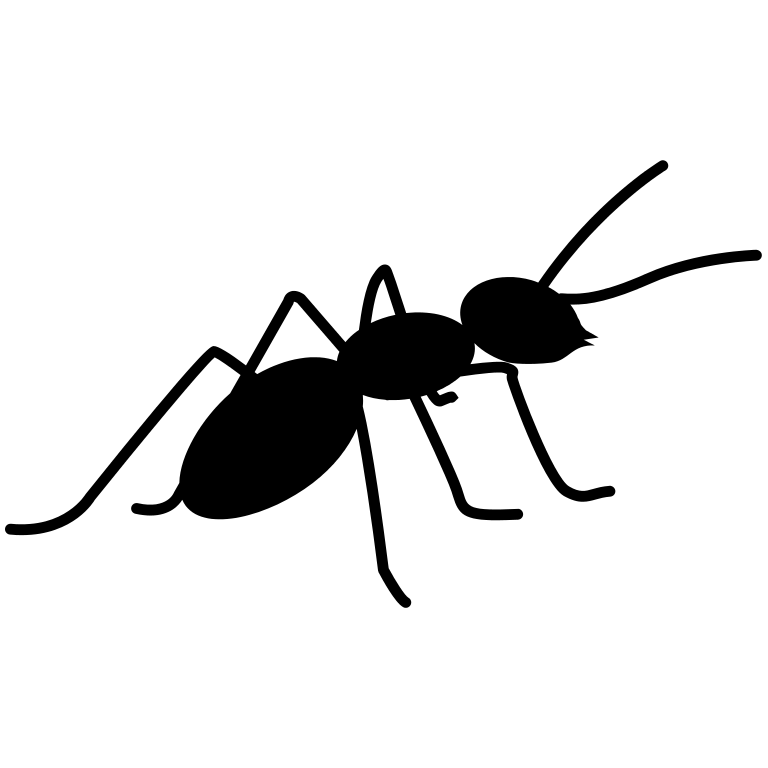}};
    
    \begin{scope}[on background layer]
    \draw [draw=black, style={decorate, decoration=snake}, thick] (start.center)--(final.center) ;
    \draw [draw=black, loosely dashdotted, thick] ([yshift=2pt]start.center)--([yshift=2pt]nodo1.center)--([yshift=2pt]final.center);
    \draw [draw=black, loosely dashed, thick] ([yshift=-2pt]start.center)--([yshift=-2pt]nodo1.center);
    \draw [draw=black, loosely dashed, thick] ([xshift=-2pt]nodo1.center)--([xshift=-2pt]nodo2.center);
    \draw [draw=black, loosely dashed, thick] ([yshift=2pt]nodo2.center)--([yshift=-1pt]final.center);
    \draw [draw=black, dotted, thick] ([yshift=-1pt]start.center)--([yshift=2pt]nodo2.center);
    \draw [draw=black, dotted, thick] ([xshift=2pt]nodo1.center)--([xshift=2pt]nodo2.center);
    \draw [draw=black, dotted, thick] ([yshift=-2pt]nodo1.center)--([yshift=-2pt]final.center);
    \draw [draw=black, thick] ([yshift=-5pt]start.center)--([yshift=-2pt]nodo2.center)--([yshift=-5pt]final.center);
    \end{scope}
\end{tikzpicture}}

\caption{Graphic representation of the paths in ACO and QACO in a 2 node graph for a path finding problem. In ACO, the ant arrives at the end by choosing a non visited node at each step. In this example, the maximum steps for the classical ant is 3, with 5 different paths. However, 2 of these paths yields the same result, since the codification only depends on the visited nodes, not in the order they are visited. In QACO, the ant lives in a state that is a superposition of each possible path. Using 2 qubits, we can assign each one to a node, where $\ket{1_i}$ and $\ket{0_i}$ represents the ant visiting and not the node $i$ respectively. In the moment the ant is measured, one of the paths is selected.}
\label{fig:antDiagram}
\end{figure*}

The main reason why we developed this algorithm is to provide a practical application of the well-known ACO algorithm on a quantum computer (Figure \ref{fig:antDiagram}). Although no real implementation is mentioned in this section, we designed it so that the steps and the gates used are easily implementable on the available computers to the date this is written.

Following an almost identical schema used in ACO, this algorithm can be divided into 4 main steps: pheromone application, exploration of new solutions, post-measurement checks and pheromone update. 

The qubits are divided into two groups: ant and exploration qubits. The ant qubits are the ones in which the information about pheromone trail is introduced, while the exploration qubits are determined by the exploration parameter.

\subsection{Algorithm step development}
\label{sec:AlgorithmStepDevelopment}

\subsubsection{Pheromone application}
\label{sec:PheromoneApplication}
The ant qubits are the targets of the controlled gates for the exploration process. The codification of the solutions must be given as a map $v$ from the solution space $S$ to the Hilbert space $\mathcal{H}$. The number of qubits used ($n$) must be sufficient so that the map $v:S\rightarrow\mathcal{H}$ can be a bimorphism, this is $v^{-1}(v(s))=s,\ \forall s\in S$. 

Contrary to the usual codification of ACO, in QACO we encode the pheromones not into the edges of a graph, but in its nodes. Then, the probability of visiting each node is given by the pheromones deposited into them. This information is introduced on the quantum state via a rotation gate. But, while the previous algorithm uses a rotation on the real plane, we apply a Y-rotation (RY) gate on each qubit. To control the possible outcomes, we limit all the rotation angles of the ant qubits to $0<\theta_i<\pi$. This way, the state of each ant qubit after we apply the rotation gate is
\begin{equation}
    \ket{\Psi_i}=R_y(\theta_i)\ket{0}=\sin\left(\theta_i/2\right)\ket{0}+\cos\left(\theta_i/2\right)\ket{1}.
\end{equation} 

We can extend this to the full quantum state for the ant qubits, just by taking the tensor product of each qubit. The result is a superposition of all the states of the Hilbert space canonical basis
\begin{equation}
    \ket{\Psi_\text{ant}}=\ket{\Psi_1}\otimes\cdots\otimes\ket{\Psi_n}=\sum_{k=0}^{2^n}\alpha_k\ket{k},
\end{equation}
where $\ket{k}$ is the state represented by the binary expansion of $k$ on $n$ bits and $\alpha_k$ is the amplitude of the corresponding state.

On the first iteration, each possible state has to have the same probability to be measured, this is $\lvert\alpha_k\rvert=1/2^n$. Therefore, the starting values of the rotation angles must be $\theta_i^0=\pi/2$. 

\subsubsection{Exploration of new solutions}
The exploration of new solutions has two aspect related to it: one related to the classical probability of exploring and the other related to how this exploration is performed.

\subsubsection*{Exploration Parameter}
The probability for exploration of new solutions can be defined by one parameter, $0\leq\beta_\text{e}\leq1$. In this case, we use the angle of a RY gate to code this parameter as the probability to measure the excited state of that qubit. On an ideal quantum computer in which the graph of connectivity between qubits is fully connected, we would only need one qubit. But most of the quantum computers currently available have at most 4 pairings for each qubit, as this is the case for the IBM \textit{Yorktown-ibmqx2}. In order to avoid the use of SWAP gates to implement the controlled gates, this qubit can be ``duplicated" simply by applying the same RY gate to different qubits. 

These qubits control the gates for the exploration process. As the probability to apply a gate depends on the probability to measure the excited state and the state of a qubit must be normalized, the exploration qubit state can be written as
\begin{equation}
    \ket{\Psi_e}=\sqrt{1-\beta_\text{e}}\ket{0}\pm\sqrt{\beta_\text{e}}\ket{1} \Rightarrow \left|\braket{1|\Psi_e}\right|^2=\beta_\text{e}.
\end{equation} 
We can generate this state using an RY gate with an angle of $\theta_e=2\arcsin(\sqrt{\beta_\text{e}})$.

To correctly implement the exploration strategy, we must reset the exploration qubit after each controlled gate is applied. This way, we avoid the entanglement between the ant and the exploration qubits. If we don't reset the qubit, the system would only have two possible outcomes, one with all controlled gates applied and the other without. We further explain this effect on appendix \ref{apx:Exploration}.

\subsubsection*{Ant exploration strategy}
In ant colony algorithms, the mechanism which allows to have new solutions is based on random exploration. The classical strategy decides to randomly explore based on a exploration parameter, that gives the frequency at which this random exploration happens. This decision is made at every step on the path, avoiding backtracking on already visited edges.

In QACO, we can introduce this mechanism explicitly using controlled gates. When the problem is unconstrained, this can be implemented using CNOT gates. This allows the system to arrive to every possible solution from any node in the graph. Using the exploration qubits as the control qubit, we can target each of the ant qubits. This way, the probability for a qubit to change its state is $\beta_\text{e}$. Applying this to every other qubit, the probability of flipping $k$ qubits is $P(k)=\beta_\text{e}^k(1-\beta_\text{e})^{n-k}$. Thus, in each iteration there is a non-zero probability for the algorithm to yield an arbitrary solution. However, the probability of exploring $k$ times decreases with $k$, thus, we favour the local exploration of solutions.

This strategy is useful when the problem is unconstrained, that is, when all the possible outcomes are a valid solution for the problem. But when there are restrictions, this strategy may result in measuring incorrect solutions. To improve the efficiency of the exploration in these cases, we propose using gates that acts on the state taking it from an allowed solution to another. This way, if we consider an ideal quantum computer, we keep the probability of measuring an allowed solutions constant. Equivalently, the leak of probability to a non allowed state is ideally 0. Then, one would have to design a strategy for the specific constraint set of the problem to solve.

In order to illustrate this concept, let us suppose that the constraint we want to address allows solutions with $m$ 1's. For this particular problem, flipping one qubit would turn a valid solution to a non valid one. However, flipping 2 qubits at the same time preserve the number of ones of the state. To apply this change to a state, we can apply a Fredkin gate. The Fredkin gate can be understood as a controlled SWAP gate between 2 target qubits. This gate maintains the probability of measuring a certain number of excited qubits. The number of gates needed to explore the whole space is $n(n-1)/2-1$, as we need one gate for each different pair of qubits and applying all possible swap gates would yield the same state as the initial.

Being the Fredkin gate with the control qubit $c$ and two targets $t_1$ and $t_2$ (CSWAP($c,t_1,t_2$)), the commutation rule between 2 Fredkin gates with the same control qubit is
\begin{equation}
    [\text{CSWAP}(c,m,n),\text{CSWAP}(c,x,y)]
    \begin{cases}
        =0 & \text{ if } \{m,n\}\cap\{x,y\}=\emptyset \text{ or } \{m,n\}=\{x,y\},\\
        \neq0 & \text{ otherwise}.
    \end{cases}
\end{equation}
As the Fredkin gates don't commute, the order in which the gates are applied determines the states the system can jump to. As there are multiple ways to explore, the order in which the gates are applied must be randomized. This way, even though the exploration is biased on each generation, the effect is averaged out through all the iterations. In addition to this, the iterative nature of the exploration process could also average out the possible errors generated while applying any of the gate throughout the process.

This exploration strategy generates a entangled state that efficiently encodes the paths of the ants as a superposition. The entanglement is made so that in an ideal quantum computer, the measurement of all the ant qubits corresponds to a path.

\subsubsection{Post-measurement checks: solution generator}
\label{sec:GenS}
In the cases when the problem is constrained to certain solutions, the measurement of the ant qubits can turn out to be an invalid solution. This invalid result comes from the information about the pheromone state or the efect of noise. Instead of discarding the solution, we propose to generate a new result.

The new solution must be as close to the previous one as possible. Following this idea, we choose to use the Hamming distance between the solutions to distribute the probabilities, so that the closer ones are favoured. We set the probability of choosing a solution as inversely proportional to the Hamming distance from the original measurement, p$_i$ $\propto$ 1/d$_i$. We define $p_i\cdot d_i = p_j\cdot d_j\ \forall i,j$. Working with both equations, and the condition that the sum of the probabilities is 1, one arrives to 
\begin{equation}
p_i^{-1} = d_i\sum_j \frac{1}{d_j}.
\end{equation}
The algorithm to choose a solution is presented on Algorithm \ref{alg:GenS}.

\begin{algorithm}[h!]    
    \caption{Algorithm to generate a new solution from an invalid measurement.}
    \label{alg:GenS}
    \begin{algorithmic}[1] 
        \Require Invalid solution b, set of all valid solutions \{s$_i$\}.
        \State Calculate the Hamming distance between b and each s$_i$, d$_i$ = d(b,s$_i$).
        \State Sum the inverse of all Hamming distances, Q = $\sum_i$ 1/d$_i$.        
        \State Calculate the probability to choose each solution, p$_i$ = 1/(Q$\cdot$d$_i$).
        \State Pick a random solution following the probability distribution given by p$_i$.
    \end{algorithmic} 
\end{algorithm}

\subsubsection{Pheromone update}
\label{sec:PheromoneUpdate}
At the end of each generation a solution is produced. If the stopping criteria is not met, we update the pheromones using a new lookup table (Table \ref{tab:PheromAngle}). This table takes into account the best solution obtained so far $\left(f(b)\right)$ and the solution for the current generation $\left(f(x)\right)$. The idea beneath this values is to implement the same mechanism used in ACO. We reinforce the best solutions by updating the rotation angle so that in the next generation the probability to measure it increases. But when a better solution is found, the rotation angle update is higher. This way, we reinforce positively the exploration of new best solutions.

To update the angle value, the angle for the next iteration for each ant qubit ($\theta_i'$) is calculated by summing the value obtained from the lookup table (Table \ref{tab:PheromAngle}) to the value used in the current iteration ($\theta_i$), $\theta_i'=\theta_i+\Delta\theta_i$. The election of values on the table is discussed on section \ref{sec:parameteropt}. 

\begin{table}[h]
    \centering    
    \caption{ Lookup table for the pheromone rotation angle update $\Delta\theta_i$. $x_i$ is the state of the qubit $i$ on the current generation, and $b_i$ the state of the qubit $i$ on the best solution so far. $f(x)$ and $f(b)$ are the values for the fitness function for the current generation and the best solution so far respectively. Values with * are multiplied by -1 if cos($\theta_i$/2)$<$0.}
    \vspace{5pt}
    \begin{tabular}{cccc}
    \hline\noalign{\smallskip}
    $x_i$ & $b_i$ & f($x$) better than f($b$)? & $\Delta\theta_i$ \\
    \noalign{\smallskip}\hline\noalign{\smallskip}
    0     & 0     & True                         & -0.01$\pi$*                          \\ 
    0     & 0     & False                        & \phantom{-}0.04$\pi$\phantom{$^*$}   \\ 
    0     & 1     & True                         & -0.05$\pi$*                          \\ 
    0     & 1     & False                        & \phantom{-}0.07$\pi$\phantom{$^*$}   \\ 
    1     & 0     & True                         & \phantom{-}0.05$\pi$*                \\ 
    1     & 0     & False                        & -0.07$\pi$\phantom{$^*$}             \\ 
    1     & 1     & True                         & \phantom{-}0.01$\pi$*                \\ 
    1     & 1     & False                        & -0.04$\pi$\phantom{$^*$}             \\
    \noalign{\smallskip}\hline
    \end{tabular}
    \label{tab:PheromAngle}
\end{table}

In this algorithm we force the values of the rotation angle to the interval [$0$, $\pi$]. If an angle is out of this interval, the next update will try to correct the angle back. When the algorithm is near convergence, the rotation angle will oscillate around 0 or $\pi$, and the state of the qubit after applying the RY gate will oscillate as well around $\ket{0}$ or $\ket{1}$.

Note that we have defined the angle update values in terms of a single ant. In ACO there is a choice of strategies for updating the pheromone trails. In this regard, QACO could benefit from exploring other update strategies, in which the pheromone rotation angles could depend on the fitness value or in more than one ant, among other possibilities. However, as it will be shown in section \ref{sec:ConvergenceOfQACO} one ant suffices for the algorithm not to converge to suboptimal solutions. This statement agrees with the results found for another hybrid quantum algorithm in \cite{StochasticGradient}.

\subsection{Stopping criteria}
In ACO, we have to define a termination condition for the algorithm to exit the iteration loop. When we have no prior information about a lower bound for the optimal solution, we can define 2 different conditions \citep[p. 105]{ACObook}. One can be to set a fixed maximum time or iterations the algorithm can run. Using this criterion, making an infinite number of iterations will yield the correct result to the problem, as every possible path is allowed to be obtained in every iteration. This way, the probability of getting the result after infinite iterations will be 100\%. Although valid, this termination criterion is not useful, as it is difficult to set the correct number of iterations a priori. Besides, the number of iterations could be set higher than necessary, lowering the efficiency of the algorithm.

The other termination condition can be set to define a convergence or stagnation condition. This can be understood as having a situation in which no better results are found on consecutive iterations. To take this into account, we introduced a new parameter \textit{converCondition}. At the end of each iteration, the algorithm checks if the result is better than the best solution so far. If this is true, the condition counter is reset to 0. Else, the counter increases in 1. When this counter is equal to \textit{converCondition} the algorithm stops and returns the best solution so far.

\subsection{Algorithm}
\begin{figure}[h]


\scalebox{0.9}{
\begin{tikzpicture}[node distance=1.5cm]
    \draw [draw=black, fill=lightgray!25!, rounded corners] (-4.6,-10) rectangle (4.6,-2.25);
    \node (quantum) at (-3.5,-3) {\normalsize \shortstack{Quantum\\processes}};
    \node (start) [startstop] {Start};
    \node (iniFerom) [process, below of=start] {Initialize pheromones};
    \node (iniAnt) [process, below of=iniFerom, fill=white] {Initialize qubits at $\ket{0}$};
    \node (Ferom) [process, below of=iniAnt, fill=white] {Apply pheromones};
    \node (tipo) [decision, below of=Ferom, yshift=-2ex, fill=white] {Problem type};
    \node (CQAPup) [coord,left of=tipo, xshift=-10ex] {};
    \node (UQAPup) [coord,right of=tipo, xshift=10ex] {};
    \node (toff) [process, below of=CQAPup, fill=white] {\shortstack{Fredkin gate\\ exploration}};
    \node (cnot) [process, below of=UQAPup, fill=white] {\shortstack{CNOT gate\\ exploration}};
    \node (CQAPdw) [coord, below of=toff] {};
    \node (UQAPdw) [coord, below of=cnot] {};
    \node (conver) [coord, below of=tipo]{};
    \node (medir) [process, below of=conver, fill=white] {Measure quantum state};
    \node (GenS) [process, below of=medir] {Apply solution filter GenS};
    \node (estadoFin) [process, below of=GenS] {Get solution};
    \node (cond) [decision, below of=estadoFin, yshift=-2ex] {¿Convergence?};
    \node (feromUpd) [process, right of=cond, xshift=13ex] {\shortstack{Update\\pheromones}};
    \node (stop) [startstop, below of=cond, yshift=-2.2ex] {Result};
    \node (auxUpd) [coord, right of=feromUpd] {};
    \draw [arrow] (start.south) -- (iniFerom);
    \draw [arrow] (iniFerom.south) -- (iniAnt);
    \draw [arrowR] (iniAnt.south) -- (Ferom);
    \draw [arrowR] (Ferom.south) -- (tipo);
    \draw [arrowR] (tipo.west) -- node[anchor=south] {Constrained\hspace*{15pt}} (CQAPup) -- (toff);
    \draw [arrowR] (tipo.east) -- node[anchor=south] {\hspace*{15pt}Unconstrained} (UQAPup) -- (cnot);
    \draw [arrowR] (toff.south) -- (CQAPdw) -- (medir);
    \draw [arrowR] (cnot.south) -- (UQAPdw) -- (medir);
    \draw [arrow] (medir.south) -- (GenS);
    \draw [arrow] (GenS.south) -- (estadoFin);
    \draw [arrow] (estadoFin.south) -- (cond);
    \draw [arrow] (cond.east) -- node[anchor=south] {No\hspace*{1.5ex}} (feromUpd);
    \draw [arrow] (cond.south) -- node[anchor=east] {Yes\hspace*{1.5ex}} (stop);    
    \draw [arrow] (feromUpd.east) -- (auxUpd) |- (iniAnt);
    \node (bordeIzq) [coord, left of=cond, xshift=-40ex] {};
    \node (bordeDer) [coord, right of=cond, xshift=40ex] {};
    \draw [-,opacity=0] (bordeIzq) -- (bordeDer); 
\end{tikzpicture}}

\caption{ Flux diagram of the proposed QACO algorithm. The steps within the box are performed in a quantum computer, and the dashed lines indicate that the information between steps is in a quantum state.}
\label{fig:QACOpropDiagrama}
\end{figure}
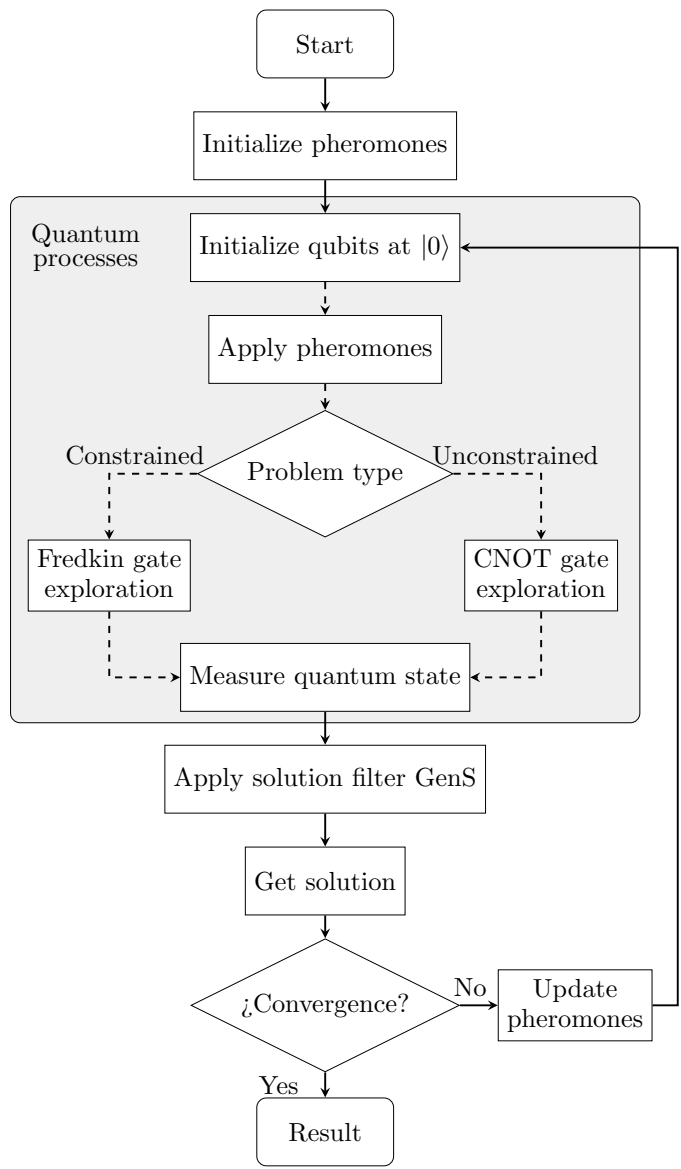

The implementation of the algorithm is very similar to the classical ACO. Having discussed the steps on the previous sections, the algorithm is presented in Algorithm \ref{alg:finQACO}. Figure \ref{fig:QACOpropDiagrama} shows a flux diagram showing the workflow of the algorithm. In Figure \ref{fig:QACOprop} we expand an example of the diagram for the quantum circuit that implements one iteration of the algorithm. 

The quantum state of the ant qubits is measured each iteration. After the post measurement checks and the pheromone update, a new quantum state is generated at the start of a new iteration. Hence, the iterative nature of the proposed algorithm.

\begin{algorithm}    
    \caption{Implemetation for QACO}
    \label{alg:finQACO}
    \begin{algorithmic}[1]
        \Require Problem $f$, pheromone update table, exploration parameter $\beta_\text{e}$,  maximum iterations $maxIter$, convergence condition $converCondition$
        \State Initialize all pheromones, $\theta_i$ = $\pi/2$.
        \For{j = 1:maxIter}
        \State Reset the quantum state of the system, $\ket{\Psi_i}=\ket{0} \forall i$.
        \State Use RY gates to apply the pheromones and the exploration parameter.        
        \State Using the exploration qubits as controls, explore with CNOT or Fredkin gates.
        \State Measure the state of the ant qubits.
        \If{the measurement is an invalid solution}
            \State Use algorithm \ref{alg:GenS} to generate a valid solution.
        \EndIf
        \If{convergence criteria is met}
            \State \bb{break}
        \EndIf
        \State Update the pheromone values.
        \EndFor
        \State \bb{return} the best solution found
    \end{algorithmic} 
\end{algorithm}

\begin{figure*}[h]


\centering
\resizebox{\textwidth}{!}{
\begin{quantikz}
    \lstick{Exploration qubit $\ket{0}$\phantom{$^{\otimes n}$}} & \gate{R_y(\theta_{e,j})} & \ctrl{2}& & \gate{R_y(\theta_{e,j})} & \ctrl{3} &  & \gate{R_y(\theta_{e,j})} & \ctrl{2}\\
    \lstick[wires=3]{Ant qubits $\ket{0}^{\otimes n}$} & \gate{R_y(\theta_{1,j})} & \swap{1} & \qw  & \qw  & \qw  & \qw & \qw  & \swap{2} & \meter{} & \cwbend{1}\\
    & \gate{R_y(\theta_{2,j})}& \targX{}  & \qw & \qw & \swap{1} & \qw  & \qw  & \qw & \meter{} & \gate[/quantikz/nwires=1]{GenS}\cw & \cw &\rstick{Solution for\\the iteration j}\\
    & \gate{R_y(\theta_{3,j})}& \qw & \qw  & \qw & \targX{} & \qw & \qw  & \targX{} & \meter{} & \cwbend{-1}
\end{quantikz}}

\caption{Example of the diagram for the circuit that implements the $j^{th}$ iteration of QACO for a constrained problem size $n=3$. The pheromone update is made once the solution for the iteration is selected. The solution checking and conditional solution generation is shorten as ``GenS".}
\label{fig:QACOprop}
\end{figure*}
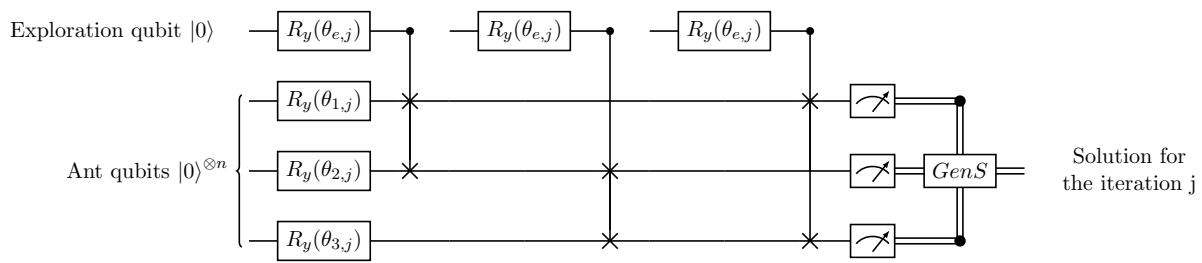

\subsection{Convergence of QACO}
\label{sec:ConvergenceOfQACO}
It is easy to see that the algorithm we propose here will arrive at the optimal solution given enough iterations, since there is a non-zero probability of measuring every possible solution at each iteration. For analyzing the convergence behaviour of QACO, let's analyze it in a worst-case scenario where the algorithm is trapped in a local minimum.

If QACO is trapped in a suboptimum point, the pheromones will initially guide the ants towards a suboptimal configuration $s$. This means that the state of the ant qubits will be close to the corresponding state of the computational basis, $\ket{\Psi_\text{ants}}=R_y\left(\theta_{k,j}\right)^{\otimes n}\ket{0}^{\otimes n}\approx\ket{s}$. In this case, the algorithm completely depends on the exploration strategy to search for a better solution. As we have shown in section \ref{sec:ExplorationStrategy}, the algorithm favours the local search of new solutions. Let's again take the worst scenario in which the ant has only searched once. Most likely, the new solution will have a worse fitness value than the local minimum. However, the fact that we have obtained a different solution introduces a variation in the pheromones. This way, the probability of searching new solutions has increased compared to the previous iteration. We can check this by calculating the projection of the state after applying the pheromones to the state encoding the suboptimal configuration, and noticing that the state of the next generation has decreased probability of being in the $\ket{s}$ state, $\left|\braket{s|\Psi_\text{ants}(k)}\right|^2>\left|\braket{s|\Psi_\text{ants}(k+1)}\right|^2$.

The likelihood of exiting a local minimum and finding a better solution is at least $\beta_\text{e}^q\left(1-\beta_\text{e}\right)^{p-q}$, with $p$ the number of different exploration operations and $q$ the number of exploration operations that separates the local and global minimum solutions. Furthermore, due to the small probability for the ant in a given exploration not to explore of $\left(1-\beta_\text{e}\right)^p$, our exploration strategy does not require more than one ant to start escaping the minimum point. Given that each time the algorithm finds a different solution the probability of searching new ones increases, it is proven that QACO will never converge to a local minimum.

\section{Implementation of QACO}
\label{sec:Implementation}
Ant Colony algorithms are usually constructed to obtain an approximate solution for NP-complete problems \citep[pp. 463-465]{DefNP}. Every NP-problem is equivalent to every other problem in the set up to a polynomial time transformation \citep{NP-hard}. This means that we can choose to solve one set of them, in this case we have chosen to solve the Quadratic Assignment Problem (QAP) \citep{QAPdef}.

In order to correctly analyse the results we have obtained, we have to keep in mind the ``No Free Lunch Theorem" \citep{NFLT}. This theorem states that a global optimization strategy does not exist over the complete set of problems. We can only obtain a better efficiency if we limit ourselves to solve a particular kind of problem.

\subsection{Quadratic Assignment Problem (QAP)}
\label{sec:QAP}
The QAP consist on searching the input $\bb{X}$ that maximizes the function given by
\begin{equation}
    f(\bb{X},\mathcal{M})=\bb{X}^t\mathcal{M}\bb{X}=\sum_{i=1}^n\sum_{j=1}^i X_i\mathcal{M}_{ij}X_j,\ \mathcal{M}_{ij}\in\mathbb{R},
\end{equation}
where $\bb{X}$ is a column vector with values 1 or 0 and $\mathcal{M}$ is the problem matrix. For simplicity, and without losing any generality, we can choose $\mathcal{M}$ to be a triangular matrix. For the problem not to be trivial, $\mathcal{M}$ has to have both positive and negative elements. The solution may be allowed to have any number of 1's in its solution or may have some constraints. We will be referring to the first case as UQAP (Unconstrained QAP) and the latest CQAP (Constrained QAP). 

For UQAP, the solution set has size $2^n$, where $n$ is the size of the matrix. Any known exact algorithm will have an exponential complexity $\mathcal{O}(2^n)$. This rapid growth in complexity limits our capacity to simulate bigger problems, in which our algorithm would come in useful. There is also a limit in the number of qubits we can simulate or to which we have access to.

\subsection{Parameter optimization}
\label{sec:parameteropt}
Before running the algorithm, we have to select the input parameters: $converCondition$, $maxIter$, and the pheromone rotation angles. For this, we have decided to search for the set of parameters that returns a good quality of solutions for QAP problems generated with uniformly random numbers. In particular, we have aimed to minimize the number of iterations before stopping needed to obtain an optimal solution. On top of this, we have added an additional constraint for the input parameters to be considered. The criterion we used for deciding if the quality is acceptable or not is to have at least 98.5\% of correct results after running the algorithm a number of times (100 runs per instance) for different problem instances (100 instances). If the the parameters obtained yield results with lower success ratio, then they are discarded. The problems employed are generated as random triangular matrices, so that the diagonal elements have the same weight as the elements outside the diagonal on the solution. To fully test QACO, we have solved both the unconstrained version of QAP (UQAP) and the constrained version (CQAP). 

For the optimization, we have used the surrogate optimization algorithm implemented in the ''Global Optimization Toolbox" in Matlab 2019b. We run the program for every distinct problem for matrices size $n=3$ to $n=7$. As the constraint of having $m$ 1's on the solution is equivalent to having $n-m$ in terms of the size of the solution set, we omitted the values of $m>\lfloor n/2\rfloor+1$. We set the maximum function evaluation parameter to 500. In the following paragraphs, we discuss the optimization of each parameter using the the results shown in Table \ref{tab:paramOptimi}.

\begin{table*}[h]
\caption{Values of the parameters in QACO that optimizes the mean number of iterations to get a solution which is incorrect at most 1.5\% of the times over 100 runs for each of the 100 randomly generated problems. $Comb$ is the number of valid solutions to each problem, $IterM$ is the mean number of iterations until convergence, $CC$ is \textit{converCondition}, $\beta_\text{e}$ is the exploration parameter, $IterF$ is the factor so that the maximum number of iterations is given by $maxIter=converCondition\cdot IterF$, and $00T$, $00F$, $01T$ and $01F$ are the values for the rotation angle update ordered as shown in Table \ref{tab:PheromAngle}. Each row shows the results for each problem size $n,m$.}
\label{tab:paramOptimi}
\vspace{3pt}
\centering
\begin{tabular}{rrrrrrrrrrr}
\hline\noalign{\smallskip}
n & m & Comb & IterM & CC & $\beta_\text{e}$ & IterF & 00T & 00F & 01T & 01F \\
\noalign{\smallskip}\hline\noalign{\smallskip}
3 & 1 & 3    & 7.00   & 7   & 0.964 & 1.000 & 0.001 & 0.090 & 0.100 & 0.005 \\
3 & 3 & 8    & 25.87  & 24  & 0.070 & 1.038 & 0.002 & 0.063 & 0.100 & 0.091 \\
4 & 1 & 4    & 12.48  & 11  & 0.211 & 1.144 & 0.006 & 0.037 & 0.052 & 0.070 \\
4 & 2 & 6    & 18.61  & 17  & 0.001 & 1.073 & 0.039 & 0.076 & 0.005 & 0.037 \\
4 & 4 & 16   & 48.56  & 45  & 0.123 & 1.067 & 0.009 & 0.036 & 0.051 & 0.066 \\
5 & 1 & 5    & 16.70  & 14  & 0.241 & 1.249 & 0.002 & 0.032 & 0.053 & 0.093 \\
5 & 2 & 10   & 34.89  & 34  & 0.793 & 1.028 & 0.057 & 0.064 & 0.067 & 0.044 \\
5 & 5 & 32   & 91.97  & 91  & 0.087 & 1.004 & 0.062 & 0.030 & 0.025 & 0.042 \\
6 & 1 & 6    & 19.83  & 19  & 0.133 & 1.020 & 0.003 & 0.029 & 0.055 & 0.077 \\
6 & 2 & 15   & 55.00  & 55  & 0.998 & 1.000 & 0.021 & 0.020 & 0.083 & 0.026 \\
6 & 3 & 20   & 63.45  & 59  & 0.001 & 1.068 & 0.024 & 0.050 & 0.022 & 0.064 \\
6 & 6 & 64   & 156.80 & 129 & 0.082 & 1.475 & 0.033 & 0.046 & 0.054 & 0.091 \\
\noalign{\smallskip}\hline
\end{tabular}
\end{table*}

\subsubsection*{Pheromone rotation angle update}
The results obtained are not sufficient to determine an optimum parameter relation. The results seems to follow a random distribution, with no correlation between parameters. As there is no clear way to set the parameters, we decided to use the median values truncated to the second decimal (Table \ref{tab:PheromAngle}).

\subsubsection*{Exploration parameter}
In the original algorithm \citep{NovelQACO}, they proposed to use a varying exploration parameter, as it is usual in other ACO algorithms. We have tried linear $\beta_\text{e}$ parameters, with positive and negative gradient and with constant values. The best result is obtained with negative gradient. However, the difference between the results are so similar, that they might be caused purely by random fluctuations. Taking this into account, we have chosen to use the positive gradient $\beta_\text{e}$ because it yielded the most consistent results and for keeping the same argument that can be made for classic ACO algorithm. At first, the ants explore the paths randomly. As the number of iterations increases, some suboptimal paths are found. To discard the suboptimal results, the increase of the exploration parameter forces the algorithm to search for new paths. In general, the exploration parameter is
\begin{equation}
    \beta_\text{e}(i)=\beta_{e0}+\frac{1-\beta_{e0}}{maxIter}i,
\end{equation}
with $\beta_{e0}$ the exploration parameter at the first iteration, $maxIter$ the maximum iteration count, and $i$ the number of the current iteration. Using this formula, the parameter is restricted to values $0\leq\beta_{e}\leq1$. The parameters chosen are the median of the different values obtained rounded up to the second decimal, $\beta_{e0}=0.13$ and $maxIter=1.05 \cdot converCondition$.

\subsubsection*{Convergence condition}
\begin{figure}[h]
    \centering
    \includegraphics[width=300pt]{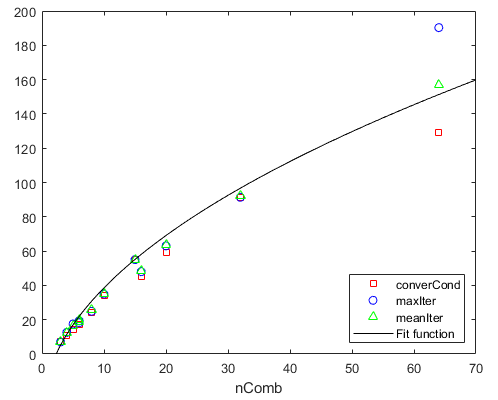}
    \caption{Results for the parameter optimization. Mean iterations until convergence $Iterm$ (circle), $converCondition$ (square) and $MaxIter$ (triangle) vs $nComb$. The solid curve is the fitting curve for $converCondition$ from equation \ref{eq:converConditionFit}.} 
    \label{fig:iterVScomb}
\end{figure}

Using the same criteria as before, we tried to optimize the \textit{converCondition} parameter. Using the same values of the Table \ref{tab:paramOptimi}, we see that the best fit value for this parameter grows as an potential function of the number of possible solutions for a problem (\textit{nComb}). Fitting the results (Figure \ref{fig:iterVScomb}) to a function dependant of \textit{nComb} we obtain
\begin{equation}
    \label{eq:converConditionFit}
    \text{converCondition}(\text{nComb})=23.3\cdot \text{nComb}^{0.5}-35.1.
\end{equation}

\subsection{Simulation of QACO}
In order to simulate these problems we have used Matlab 2019b. We have simulated an ideal quantum computer, in which there is no noise and the gates are also ideal. We can use the matrix representation for the gates and column vectors in the computational basis for the state. As every gate of the circuit is unitary, we can apply the gates in braket notation maintaining the normalization condition, $U_{gate}\ket{\Psi}=\ket{\Psi'}$, where $\ket{\Psi}$ is the state before and $\ket{\Psi'}$ after we apply the gate. As we have previously mentioned, as not all the gates commute with each other, the order in with we apply them affects the result.

The simulation of the measurement is implemented by choosing randomly a final state of the computational basis. For this, we use the probability distribution given by $|\braket{\Psi|\Psi}|$. Apart from these particularities, the algorithm follows the same steps as explained in section \ref{sec:ImplementableQACO}.

\subsection{Experiment on IBM's quantum computers}
\label{sec:ExperimentQC}
For implementing our algorithm on a real quantum computer, we have chosen to use IBM's computers. Our main goal with this implementation is to minimize the number of quantum gates needed in each iteration. As the decoherence time is still a constraint, a smaller set of gates would help to maintain the information on the system with as less perturbations as possible. For this, we have to correctly analyze the topology of the computer we will use.

\begin{figure}[h]


\makebox[\linewidth][c]{
\begin{tikzpicture}[node distance=1cm]
    \node (2) [draw=black, circle, minimum size=20pt, fill=cyan] {2};
    \node (0) [draw, circle, minimum size=20pt, left of=2] {0};
    \node (1) [draw, circle, minimum size=20pt, above of=2] {1};
    \node (3) [draw, circle, minimum size=20pt, right of=2] {3};
    \node (4) [draw, circle, minimum size=20pt, below of=2] {4};
    \draw (0) -- (1);
    \draw[thick,double] (1) -- (2);
    \draw[thick,double] (2) -- (3);
    \draw (3) -- (4);
    \draw[thick,double] (4) -- (2);
    \draw[thick,double] (2) -- (0);
\end{tikzpicture}}

\vspace*{10pt}
\makebox[\linewidth][c]{
\begin{tikzpicture}[node distance=1cm]
    \node (0) [draw=black, circle, minimum size=20pt, fill=cyan] {0};
    \node (1) [draw, circle, minimum size=20pt, right of=0] {1};
    \node (2) [draw, circle, minimum size=20pt, right of=1] {2};
    \node (3) [draw, circle, minimum size=20pt, right of=2] {3};
    \node (4) [draw=black, circle, minimum size=20pt, fill=cyan, right of=3] {4};
    \node (5) [draw, circle, minimum size=20pt, right of=4] {5};
    \node (6) [draw, circle, minimum size=20pt, right of=5] {6};
    \node (14) [draw, circle, minimum size=20pt, below of=0] {14};
    \node (13) [draw, circle, minimum size=20pt, right of=14] {13};
    \node (12) [draw=black, circle, minimum size=20pt, fill=cyan, right of=13] {12};
    \node (11) [draw, circle, minimum size=20pt, right of=12] {11};
    \node (10) [draw, circle, minimum size=20pt, right of=11] {10};
    \node (9) [draw, circle, minimum size=20pt, right of=10] {9};
    \node (8) [draw=black, circle, minimum size=20pt, fill=cyan, right of=9] {8};
    \node (7) [draw, circle, minimum size=20pt, right of=8] {7};
    \draw[thick,double] (14) -- (0) -- (1);
    \draw[thick,double] (13) -- (12) -- (11);
    \draw[thick,double] (9) -- (8) -- (7);
    \draw[thick,double] (3) -- (4) -- (5);
    \draw[thick,double] (12) -- (2);
    \draw[thick,double] (4) -- (10);
    \draw[thick,double] (8) -- (6);
    \draw (14) -- (13) -- (1) -- (2) -- (3) -- (11) -- (10) -- (9) -- (5) -- (6);
\end{tikzpicture}}

\caption{ ibmq\_5\_yorktown - ibmqx2 (above) and ibmq\_16\_melbourne (below) qubit arrangements. For QACO implementation, the optimal position of the exploration qubits for UQAP problems are colored. Double lines represent the couplings between the ant and exploration qubits, single lines represent unused couplings.}
\label{fig:ibmq}
\end{figure}
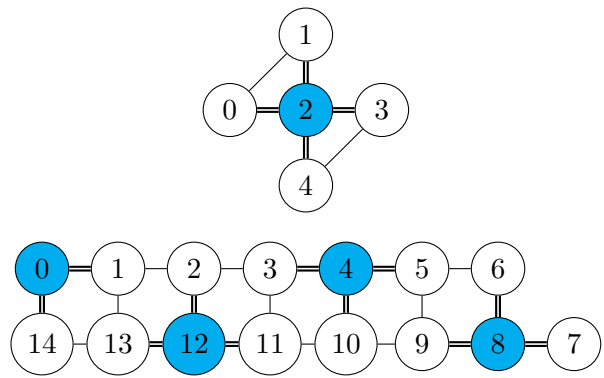

IBM's quantum computers are based on superconducting systems that have their qubits on a 2D lattice, we have a plane graph representing the possible coupling between qubits, being the nodes the qubits and the vertex the couplings. Different computers have different configurations, some of which have a connectivity graph more suited for this algorithm. For implementing QACO we have chosen the 2 with the configurations that maximizes the number of ant qubits per exploration qubit (Figure \ref{fig:ibmq}): \textit{ibmq\_5\_yorktown - ibmqx2} and \textit{ibmq\_16\_melbourne}.  

Depending of the problem size and type (constrained or unconstrained) one has to find the correct computer and arrangement of qubits. In the case of unconstrained problems, the aim is to maximize the number of ant qubits while having them connected to at least one exploration qubit. This way, we can apply CNOT gates without having to use SWAP gates that introduce errors on the system. As we can see in Figure \ref{fig:ibmq}, square lattices might be the best configurations to solve this problems. In the current generation superconducting circuits, square lattices let us connect up to 3 ant qubit to a single exploration qubit. The downside of this arrangement is that we lose some of the entanglement of the system. This could lead to a worse convergence velocity, as the final measurement is allowed to be a combination of different paths. A solution for this problem could be to assign each qubit to a random position in the system in each iteration. Similarly to the exploration strategy with Fredkin gates, this defects could be averaged over all iterations.

If we have constrained problems, we need more connectivity between qubits. As we need to apply Fredkin gates, we need 3 qubit cycles in the connectivity graph. This high connectivity is a problem in large scale superconducting quantum circuits. At the moment, the only IBM computer that has this type of topology is the \textit{ibmq\_5\_yorktown - ibmqx2} computer. If the connection graph for this computer were complete we could solve any type of unconstrained problem with size $n=4$. Unfortunately we only have 2 cycles in this graph, so we are limited to solve constrained problems in which the solution is restricted to have one 1 in two subsets of two qubits. 

On top of these problems, we are limited to the constraints of the IBM provider. As our algorithm changes the gate's parameters at each iteration, we can not run the algorithm at once. This forces us to send new petitions to the provider sequentially. Taking these limitations into account, we have run the algorithm on the IBM quantum computers just once, in order to check that the algorithm successfully converges to the optimal configuration for each problem instance.  

\subsection{ACO}
We have also used ACO to solve the problems in order to have a fair comparison between QACO and its classical counterpart. For this we have used a simple version of ACO, which is based on the original article from \cite{HCF}. This version is summarized into algorithm \ref{alg:ACO}. The input parameters for the algorithm are similar to the ones used for QACO, with the same $\beta_\text{e}=0.13$, $maxIter=62$ and $converCondition=59$, while the pheromone evaporation value $\rho=0.05$ is the same as the one in the original ACO paper. To fully mimic the implementation of QACO, we have only launched one ant per iteration of ACO. However, a key point of ACO is to have a swarm of ants exploring new solutions. Thus, we have also allowed the algorithm to have more than one ant per iteration.

\begin{algorithm}
    \caption{ACO}
    \label{alg:ACO}
    \begin{algorithmic}[1]
        \Require Problem matrix $\mathcal{M}$ of size $n\times n$, pheromone evaporation $\rho$, exploration parameter $\beta_\text{e}$,  maximum iterations $maxIter$, convergence condition $converCondition$
        \State Initialize all pheromones, $\tau_{i,j}$ = 0, $i,j=0,\dots,n+1$.
        \State Initialize the iteration count, count $=1$
        \For{j = 1:maxIter}
            \State Initialize the position of the ant at position $p=0$
            \For{k = 1:size of $\mathcal{M}$}
                \If{random number $\in (0,1)<\beta_\text{e}$}
                    \State Move to a random non visited position 
                \Else
                    \State Move to a random non visited position according to the pheromone trail with probability, $P(i)=\tau_{p,i}/\sum_{\ell=non\ visited\ pos.}\tau_{p,\ell}$
                \EndIf
                \If{ant is at position $n+1$}
                    \State Break the loop
                \EndIf
            \EndFor
            \State Generate a vector where the corresponding visited positions are marked with 1, $X$.
            \State Calculate the fitness of the result, $F(X)=X\mathcal{M}X'$
            \State Update de pheromones according to the path traveled $\tau_{i,j}^{(ant)}$, $\tau_{i,j}^{(j+1)}=(1-\rho)\tau_{i,j}^{(j)}+(\rho/F(X))\tau_{i,j}^{(ant)}$
            \If{$F(X)$ is the best solution found so far}
                \State $count = count +1$
            \EndIf
            \If{$count = converCondition$}
                \State Break the loop
            \EndIf
        \EndFor
        \State \bb{return} the best solution found
    \end{algorithmic} 
\end{algorithm}

\subsection{Results}
We have successfully computed a small set of unconstrained problems. The benchmark instances that are more often used to test the performance of algorithm solving QAP are too large to be solved by our simulator. The typical instance sizes go from 25 \cite{glover} to 7000 variables \cite{Palubeckis}. As we have limitations on the size of the problems we can simulate due to memory consumption of quantum simulations and limitations on the number of qubits on the free access IBM's computers, we have decided to solve smaller problems. Since, as to our knowledge, there are no benchmark instances of this size, the problem instances have been generated randomly as triangular matrices of different sizes. In particular, we solved 5 problems size $n=4$ on \textit{ibmq\_5\_yorktown - ibmqx2} computer.\footnote{When we started this paper, the \textit{ibmq\_16\_melbourne} computer was still online.  However, at the moment of running the algorithm it is scheduled to be retired. Although we had planned to solve an $n=11$ sized problem, we couldn't run it in time.} The problem instances used are written in Appendix \ref{apx:Probs}.

We have compared the results obtained from IBM computers with simulations of QACO and ACO we made in Matlab. For this, we have computed $100$ different runs of the algorithms. For ACO we have first launched the algorithm with just one ant per generations. The performance of each algorithm is measured by two parameters obtained after running the algorithm a number of times for the same problem. For testing the convergence rate of the algorithm we have used the mean number of iterations for the algorithm to stop. For testing the quality of the results we have measured the percentage of incorrect solutions, which was previously obtained using an exact algorithm. In order to have a fairer comparison between ACO and QACO, we have also run ACO with more than one ant per iteration. The objective of this fair comparison is to obtain the same quality solutions. To achieve this, we have run ACO 100 times for the same problem with one ant per iteration. If there were more than 1\% of incorrect solutions, we rerun the algorithm for one extra ant, until the results met our criteria. On Table \ref{tab:results} we show the outcomes of the two performance parameters for 5 problems of size 4. The experiments done in the IBM quantum computers produced the correct answer in every trial we made.

Although not tested, launching the experiment a number of times could reproduce the results obtained in the simulations, with little to no differences. More importantly, it is shown that QACO outperforms ACO in terms of consistency of finding the optimal solution to the problems with the same number of ants per iterations. Since the size of the problems we tested are small, the algorithms can only stop at a small set of different iteration numbers. This makes the probability distribution of the exit iteration number narrow, thus, we can not extract any conclusion about the convergence speed.

\begin{table}[h]
\centering
\caption{Results of the benchmark given by the problem matrices from the appendix \ref{apx:Probs}. For ACO and the simulation of QACO, it is shown the mean number of iterations before exiting the algorithm and the percentage of errors committed over 100 runs of the algorithm for each problem. We also show the number of ants per generation used in ACO. QACO has only been run once per problem on the IBM's computers. In this case, we show the iteration at which the algorithm exited, and a error \% of 0, showing that the algorithm reached the optimal solution.}
\begin{tabular}{cl|r|r|r|r|}
\cline{3-6}
\multicolumn{1}{l}{}                                   &           & \multicolumn{1}{c|}{QACO IBM}   & \multicolumn{1}{c|}{QACO}       & \multicolumn{2}{c|}{ACO} \\ 
\multicolumn{1}{l}{}                                   &           & \multicolumn{1}{c|}{single run} & \multicolumn{1}{c|}{simulation} & \multicolumn{2}{c|}{}    \\ \hline
\multicolumn{1}{|c|}{\multirow{3}{*}{$\mathcal{M}_1$}} & Mean iter & 60                              & 60.62                           & 60.22     & 60.73        \\
\multicolumn{1}{|c|}{}                                 & Ants      &                                 &                                 & 1         & 4            \\
\multicolumn{1}{|c|}{}                                 & Error \%  & 0                               & 0                               & 53        & 0            \\ \hline
\multicolumn{1}{|c|}{\multirow{3}{*}{$\mathcal{M}_2$}} & Mean iter & 58                              & 60.69                           & 60.67     & 60.49        \\
\multicolumn{1}{|c|}{}                                 & Ants      &                                 &                                 & 1         & 8            \\
\multicolumn{1}{|c|}{}                                 & Error \%  & 0                               & 1                               & 62        & 0            \\ \hline
\multicolumn{1}{|c|}{\multirow{3}{*}{$\mathcal{M}_3$}} & Mean iter & 61                              & 60.68                           & 59.75     & 60.50        \\
\multicolumn{1}{|c|}{}                                 & Ants      &                                 &                                 & 1         & 4            \\
\multicolumn{1}{|c|}{}                                 & Error \%  & 0                               & 1                               & 70        & 0            \\ \hline
\multicolumn{1}{|c|}{\multirow{3}{*}{$\mathcal{M}_4$}} & Mean iter & 61                              & 60.72                           & 60.70     & 60.45        \\
\multicolumn{1}{|c|}{}                                 & Ants      &                                 &                                 & 1         & 8            \\
\multicolumn{1}{|c|}{}                                 & Error \%  & 0                               & 3                               & 53        & 0            \\ \hline
\multicolumn{1}{|c|}{\multirow{3}{*}{$\mathcal{M}_5$}} & Mean iter & 61                              & 60.62                           & 60.70     & 60.74        \\
\multicolumn{1}{|c|}{}                                 & Ants      &                                 &                                 & 1         & 2            \\
\multicolumn{1}{|c|}{}                                 & Error \%  & 0                               & 0                               & 2         & 0            \\ \hline
\end{tabular}
\label{tab:results}
\end{table}

\section{Conclusions and future work}
\label{sec:Conclusion}
In this work we presented a new global search algorithm inspired on the classic ACO algorithm. The new proposed QACO is an iterative quantum hybrid algorithm that can be implemented in computers with non-error corrected qubits. Based on previous works, we use a pheromone representation on the Bloch sphere. We propose a general exploration strategy using controlled gates, which efficiently explores for new solutions in constrained problems. However, there are still some questions open for a future research, for instance, allowing a more complex pheromone update strategy or increasing the number of ants per iteration.

We have simulated the algorithm for problems sizes $n=3$ to $n=6$ to obtain the optimal parameters for random BQP problems, showing that the algorithm is capable of solving QAP optimization problems. An improvement in the simulations could be made if instead of using the vector representation for the quantum state, the density matrix representation was used. For this we would need to have more powerful computers with a higher memory capacity. This would better address the usefulness of the entanglement for the CQAP, which can not be fully simulated with the vector representation.

We give some guidelines to implement the algorithm in a quantum computer. We have in fact implemented the algorithm on a IBM quantum computer and successfully obtained the expected result results. The results of a bechmarking for a set of 5 problems of size $n=4$ shows that our QACO algorithm outperforms a simple version of ACO. However, the experiments done are not sufficient to fully prove the usefulness of this algorithm. In regard to the implementation on a real quantum computer, we are confident that if we set a experiment with a larger number of trials, the results will match the simulations. We also expect that for larger problem sizes, the proposed QACO algorithm could outperform ACO in terms of obtaining the optimal result in less iterations.

\begin{appendices}
\renewcommand{\thesection}{1}
\section{Problem matrices}
\label{apx:Probs}
$\mathcal{M}$ are the problem matrices and $x$ the solution:
\begin{equation*}
    \mathcal{M}_{1}^{(4)}=\begin{pmatrix}
    -0.269 & 0.411 & -0.079 & 0.175 \\
    & -0.086 &-0.222 & -0.170 \\
    & & -0.463 & 0.244 \\
    & & & -0.139
    \end{pmatrix},\ x_{1}^{(4)}=[1100],
\end{equation*}\vspace*{0.01pt}
\begin{equation*}
    \mathcal{M}_{2}^{(4)}=\begin{pmatrix}
    0.430 & -0.496 & -0.443 & 0.223 \\
    & 0.254 & 0.029 & -0.359 \\
    & & -0.424 & -0.183 \\
    & & & 0.301 
    \end{pmatrix},\ x_{2}^{(4)}=[1001],
\end{equation*}

\begin{equation*}
    \mathcal{M}_{3}^{(4)}=\begin{pmatrix}
    -0.039 & -0.327 & 0.311 & 0.100 \\
    & 0.364 & 0.051 & -0.387 \\
    & & 0.271 & 0.116 \\
    & & & 0.261 
    \end{pmatrix},\ x_{3}^{(4)}=[1011],
\end{equation*}

\begin{equation*}
    \mathcal{M}_{4}^{(4)}=\begin{pmatrix}
    -0.092 & -0.425 & 0.001 &-0.116 \\
    & 0.167 &-0.110 &-0.370 \\
    & & 0.394 &-0.061 \\
    & & & 0.104 
    \end{pmatrix},\ x_{4}^{(4)}=[0110],
\end{equation*}

\begin{equation*}
    \mathcal{M}_{5}^{(4)}=\begin{pmatrix}
    0.409 & -0.195 &-0.248 & 0.132 \\
    & -0.200 & 0.242 & -0.408 \\
    & & -0.205 & 0.248 \\
    & & &-0.298
    \end{pmatrix},\ x_{5}^{(4)}=[1000].
\end{equation*}    

\renewcommand{\thesection}{2}
\section{Exploration qubit reset}
\label{apx:Exploration}

In section \ref{sec:ExplorationStrategy}, we state that in order to correctly explore new solutions, the control qubits holding the information about the exploration probability $\beta_\text{e}$ must be reset after each controlled operation. In this appendix, we illustrate this effect on a small system, but the results here can be straightforwardly extended to any system in which we apply controlled operations with a single control qubit.

\begin{figure}[h]
    \centering
    \begin{quantikz}[slice all]
        & \gate{R_y}  & \ctrl{1} & \ctrl{2} & \qw &     \\
        & \qw         & \targ{}  & \qw      & \qw & \qw \\
        & \qw         & \qw      & \targ{}  & \qw & \qw 
    \end{quantikz}\hspace{50pt}
    \begin{quantikz}[slice all]
        & \gate{R_y}  & \ctrl{1} & \qw &     & \gate{R_y} & \ctrl{2} & \qw &     \\
        & \qw         & \targ{}  & \qw & \qw & \qw        & \qw      & \qw & \qw \\
        & \qw         & \qw      & \qw & \qw & \qw        & \targ{}  & \qw & \qw 
    \end{quantikz}
    \caption{In the circuit on the left, the exploration qubit is not reset after it is employed as the control qubit for the CNOT gates. This entangles the circuit in such way that the exploration done is not properly done. In the circuit on the right, as the exploration qubit is reseted after each CNOT gate, we recover the proper exploration probabilities. The horizontal lines marks each of the circuit steps.}
    \label{fig:explorComparation}
\end{figure}
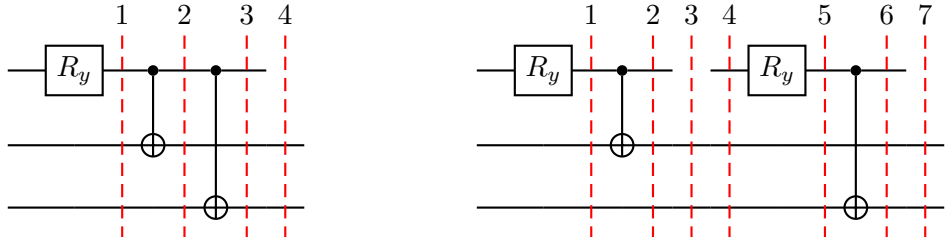

Let's suppose a system where we have a single exploration qubit and two ant qubits. The initialization of the ant qubits do not have any effect on this process, so for simplicity, suppose that the system starts with all qubits in the ground state, $\ket{\Psi_\text{in}}=\ket{0}\otimes\ket{00}$. Furthermore, suppose that the exploration strategy employed consists of CNOT gates. The two different circuits we have to compare are shown in figure \ref{fig:explorComparation}. The evolution of the quantum state for the circuit in which we do not reset the exploration qubit is the following:
\begin{equation*}
\begin{split}
    &\ket{\Psi_\text{in}}=\ket{0}\otimes\ket{00}=\ket{000} \rightarrow \text{R$_y$ on qubit 1},\\
    &\ket{\Psi_1}=\left(\sqrt{1-\beta}\ket{0}+\sqrt{\beta}\ket{1}\right)\otimes\ket{00}=\sqrt{1-\beta}\ket{000}+\sqrt{\beta}\ket{100}\rightarrow \text{CNOT 1-2},\\
    &\ket{\Psi_2}=\sqrt{1-\beta}\ket{000}+\sqrt{\beta}\ket{110} \rightarrow \text{CNOT 1-3},\\
    &\ket{\Psi_3}=\sqrt{1-\beta}\ket{000}+\sqrt{\beta}\ket{111} \rightarrow \text{Trace out qubit 1},\\
    &\ket{\Psi_\text{out}}=\sqrt{1-\beta}\ket{00}+\sqrt{\beta}\ket{11}.\\
\end{split}
\end{equation*}
For the circuit in which the qubit is reset the evolution is:
\begin{equation*}
\begin{split}
    &\ket{\Psi_\text{in}}=\ket{0}\otimes\ket{00}=\ket{000} \rightarrow \text{R$_y$ on qubit 1},\\
    &\ket{\Psi_1}=\left(\sqrt{1-\beta}\ket{0}+\sqrt{\beta}\ket{1}\right)\otimes\ket{00}=\sqrt{1-\beta}\ket{000}+\sqrt{\beta}\ket{100} \rightarrow \text{CNOT 1-2},\\
    &\ket{\Psi_2}=\sqrt{1-\beta}\ket{000}+\sqrt{\beta}\ket{110} \rightarrow \text{Trace out qubit 1},\\
    &\ket{\Psi_3}=\sqrt{1-\beta}\ket{00}+\sqrt{\beta}\ket{10} \rightarrow \text{Add qubit 1},\\
    &\ket{\Psi_4}=\ket{0}\otimes\left(\sqrt{1-\beta}\ket{00}+\sqrt{\beta}\ket{10}\right) \rightarrow \text{R$_y$ on qubit 1},\\
    &\ket{\Psi_5}=\left(\sqrt{1-\beta}\ket{0}+\sqrt{\beta}\ket{1}\right)\otimes\left(\sqrt{1-\beta}\ket{00}+\sqrt{\beta}\ket{10}\right)\\
    &\phantom{\ket{\Psi_5}}\hspace{2pt}=(1-\beta)\ket{000}+\sqrt{\beta(1-\beta)}\ket{100}+\sqrt{(1-\beta)\beta}\ket{010}+\beta\ket{110}\rightarrow \text{CNOT 1-3},\\
    &\ket{\Psi_6}=(1-\beta)\ket{000}+\sqrt{\beta(1-\beta)}\ket{101}+\sqrt{(1-\beta)\beta}\ket{010}+\beta\ket{111}\rightarrow \text{Trace out qubit 1},\\
    &\ket{\Psi_\text{out}}=(1-\beta)\ket{00}+\sqrt{\beta(1-\beta)}\ket{01}+\sqrt{(1-\beta)\beta}\ket{10}+\beta\ket{11}.
\end{split}
\end{equation*}

As we see here, in the case in which we do not reset the exploration qubit, there are only two possible outcomes: no exploration happened at all, or all possible exploration steps happened. On the contrary, resetting the exploration qubit allows for the expected exploration. In this case, the probability of applying one exploration step has the expected probability of $\beta$. This way we can achieve a smooth exploration over the whole search space, as it is intended for both ACO and QACO.

\end{appendices}

\section*{}
\textbf{Funding} The authors acknowledge support from Tecnalia and the University of the Basque Country (UPV-EHU), grant PIFTEC21/04.

\bibliographystyle{spbasic}      
\bibliography{Bibliography.bib}   

\end{document}